\documentclass[conference]{IEEEtran}
\IEEEoverridecommandlockouts

\usepackage{amsmath,amssymb,amsfonts}
\usepackage{graphicx}
\usepackage{xcolor}
\usepackage{booktabs}
\usepackage{multirow}
\usepackage{tabularx}
\usepackage{enumitem}
\usepackage{algorithm}
\usepackage[noend]{algpseudocode}
\usepackage{comment}
\usepackage{listings}
\usepackage{tikz}
\usepackage{fancyvrb}
\usepackage{pifont}
\usepackage[normalem]{ulem}  
\usepackage{caption}
\usepackage{subcaption}
\usepackage{float}
\usepackage[most]{tcolorbox}
\usepackage{url}
\usepackage{hyperref}
\hypersetup{
  colorlinks=true,
  linkcolor=black,
  citecolor=black,
  urlcolor=blue,
  pdftitle={VeriChat: An Agentic Conversational AI Assistant for Hardware Security Verification},
  pdfauthor={Dipayan Saha, Khan Thamid Hasan, Shams Tarek, Sujan Kumar Saha, Mark Tehranipoor, Farimah Farahmandi}
}
\usetikzlibrary{positioning, arrows.meta, calc}


\renewcommand{\IEEEkeywordsname}{Keywords}

\definecolor{ashgray}{RGB}{120,120,120}

\newcommand{\cmarknum}[1]{\textcolor{black}{\ding{#1}}}
\newcommand{\new}[1]{#1}
\newcommand{\hrefcite}[2]{\href{#1}{link}}

\definecolor{verilogcommentcolor}{RGB}{104,180,104}
\definecolor{verilogkeywordcolor}{RGB}{49,49,255}
\definecolor{verilogsystemcolor}{RGB}{128,0,255}
\definecolor{verilognumbercolor}{RGB}{255,143,102}
\definecolor{verilogstringcolor}{RGB}{160,160,160}
\definecolor{verilogdefinecolor}{RGB}{128,64,0}
\definecolor{verilogoperatorcolor}{RGB}{0,0,128}
\definecolor{codebg}{rgb}{0.98,0.98,0.98}
\definecolor{codegray}{rgb}{0.5,0.5,0.5}
\definecolor{codegreen}{rgb}{0,0.6,0}
\definecolor{codeblue}{rgb}{0,0,1}

\lstdefinestyle{mystyle}{
    basicstyle=\ttfamily\footnotesize,
    commentstyle=\color{codegreen},
    keywordstyle=\color{codeblue},
    showstringspaces=false,
    breaklines=true,
    frame=single,
    captionpos=b,
    tabsize=2,
    numbers=none,
    literate={**}{{\textbf{}}{}}2
}

\lstdefinestyle{prettyverilog}{
   language           = Verilog,
   backgroundcolor    = \color{codebg},
   columns            = fullflexible,
   commentstyle       = \tiny\color{verilogcommentcolor}\itshape,
   keywordstyle       = \tiny\bfseries\color{verilogkeywordcolor},
   stringstyle        = \tiny\color{verilogstringcolor},
   numberstyle=\tiny\color{black},
   identifierstyle    = \tiny\color{black},
   basicstyle         = \tiny\ttfamily,
   numbers            = none,
   showstringspaces   = false,
   frame              = single,
   tabsize            = 2,
   escapeinside       = {(*@}{@*)},
   morecomment        = [l]{//}
}

\let\OLDthebibliography\thebibliography
\renewcommand\thebibliography[1]{%
  \OLDthebibliography{#1}%
  \setlength{\itemsep}{0pt}%
  \setlength{\parsep}{0pt}%
  \setlength{\parskip}{0pt}%
  \fontsize{7.4}{7.5}\selectfont}

\begin{document}

\title{VeriChat: An Agentic Conversational AI Assistant for Hardware Security Verification}

\author{\IEEEauthorblockN{Dipayan Saha, Khan Thamid Hasan, Shams Tarek, Sujan Kumar Saha, Mark Tehranipoor, and Farimah Farahmandi}
\IEEEauthorblockA{\textit{Department of Electrical and Computer Engineering, University of Florida}\\
Gainesville, FL, USA\\
\{dsaha, khanthamidhasan, shams.tarek, sujansaha\}@ufl.edu, \{tehranipoor, farimah\}@ece.ufl.edu}
\thanks{979-8-3195-0489-0/26/\$31.00~\copyright2026 IEEE}
\thanks{We thank the U.S. National Science Foundation (NSF) for support through CAREER Award No. 2339971.}
}
\maketitle
\thispagestyle{empty}

\begin{tikzpicture}[remember picture, overlay]
  \node[anchor=north, font=\footnotesize, align=center, text width=\textwidth] at ($(current page.north)+(0in,-0.30in)$) {This paper will be presented at the 2026 IEEE International Conference on Omni-layer Intelligent Systems (COINS 2026), \url{https://coinsconf.com/}.};
\end{tikzpicture}%
\begin{tikzpicture}[remember picture, overlay]
  \node[anchor=north east, font=\normalsize] at ($(current page.north east)+(-0.62in,-0.55in)$) {(Special Session)};
\end{tikzpicture}

\begin{abstract}
Hardware security verification is a multi-stage process in which engineers must navigate complex design analyses, threat considerations, and verification strategies. They often need security-focused guidance, yet current verification environments provide little structured support for such assistance. Although conversational AI could offer such on-demand assistance, directly using general-purpose chatbots like ChatGPT or Gemini is risky due to their tendency to hallucinate and their reliance on static, outdated knowledge. We present \textit{VeriChat}, a domain-specialized conversational assistant designed to support, rather than replace, existing verification workflows by providing context-aware security guidance. \textit{VeriChat} employs a retrieval-augmented, multi-agent workflow in which three specialized agents collaboratively minimize hallucinations while improving the transparency and reliability of the response. \new{Beyond question answering, \textit{VeriChat} integrates open-source EDA tools, including Icarus Verilog, Yosys, and SymbiYosys, to perform syntax checking, synthesis analysis, simulation, and formal verification directly on user-provided RTL designs.} Evaluated using a comprehensive methodology, VeriChat achieves a Faithfulness score of 87.73\%, significantly outperforming the leading proprietary models. \new{We demonstrate the framework through a hardware Trojan detection case study on an AES S-Box IP, where \textit{VeriChat} autonomously identifies, simulates, and formally proves a covert key-leakage vulnerability through a multi-turn conversational workflow.}
\end{abstract}


\section{Introduction}

Hardware security verification is an incredibly demanding and tedious process that requires immense time and effort. Security verification engineers frequently encounter decision points throughout the verification lifecycle where they must exercise complex security-specific judgments. At many of these points, they would benefit from a helping hand or security-focused guidance. At any stage, they may need to seek \cmarknum{182} \emph{conceptual clarification and reasoning-based assessments}, for example, to understand security-relevant design behaviors, interpret threat relevance, or evaluate the soundness of their security properties and assumptions. As they move into planning and execution, they may seek \cmarknum{183} \emph{suggestions on appropriate verification methods}, aiming to identify which approaches are the most suitable for addressing specific security concerns. They may also participate in the \cmarknum{184} \emph{development of ideas oriented to security}, devising new strategies, rules, or metrics to fill the gaps left by existing approaches. During active verification, they often perform \cmarknum{185} \emph{diagnostic and debugging reasoning} to interpret ambiguous or unexpected outcomes and resolve failed checks. Finally, when results are obtained, they must perform \cmarknum{186} \emph{results validation} to determine whether the security coverage achieved is sufficient and aligned with accepted practices. Throughout these stages, engineers are repeatedly required to pause, reflect, and seek security-focused guidance, but current verification environments provide little structured support to help them do so. Throughout these stages, engineers frequently need security-focused guidance, yet current verification environments provide little structured support, forcing them to rely on manual information gathering that can lead to inefficiency, inconsistency, and overlooked vulnerabilities. This gap gives rise to our first research question:

\textbf{\emph{\underline{RQ1}}: How can security verification engineers be provided with on-demand, security-focused guidance that reduces their reliance on manual information gathering?}
\begin{figure}[!t]
\centering
\begin{tcolorbox}[
  width=\columnwidth,
  colback=blue!5!white,
  colframe=blue!60!black,
  sharp corners,
  boxrule=1pt,
  left=2pt,right=2pt,top=2pt,bottom=2pt,
  title=\textbf{Query}
]
\scriptsize
I have a Register-Transfer Level (RTL) implementation of an AES-128 core written in Verilog.
I need to perform a comprehensive pre-silicon security assessment to identify vulnerabilities to
power side-channel attacks. Specifically, I am looking for a detailed overview of established
frameworks that can be used for power side-channel leakage assessment.
\end{tcolorbox}
\begin{tcolorbox}[
  width=\columnwidth,
  colback=green!5!white,
  colframe=green!50!black,
  sharp corners,
  boxrule=1pt,
  left=2pt,right=2pt,top=2pt,bottom=2pt,
  title=\textbf{Response}
]
\scriptsize
Established Frameworks for Power Side-Channel Leakage Assessment (Pre-Silicon)

\textcolor{red}{\textbf{1. SILVER} -- \textbf{Key Idea:} Estimates power side-channel leakage from RTL via switching activity and statistical metrics (KL divergence, TVD).
\textbf{Paper:} ``SILVER: Simulation Leakage Evaluation for Hardware Security Validation'' -- HOST 2019}

\textbf{2. RTL-PSC} -- \textbf{Type:} Simulation-based + \textcolor{red}{Information-theoretic}
\textcolor{red}{\textbf{Key Idea:} Characterizes RTL designs using MI} and KL Divergence over \textcolor{red}{Hamming weight or toggle traces.}
\textbf{Features:} Metrics: \textcolor{red}{MI}, KL, \textcolor{red}{TVD}; \textcolor{red}{Trace models: HW, HD,} toggle count
\textcolor{red}{\textbf{Paper:} ``RTL-PSC: Automated Power Side-channel Leakage Assessment at RTL'' -- DATE 2020}

\textcolor{red}{\textbf{3. SCARF} -- \textbf{Key Idea:} Uses symbolic execution on RTL to identify power side-channel vulnerabilities before synthesis.
\textbf{Reference:} ``SCARF: RTL-based Side Channel Analysis using Formal Methods'' -- DAC 2018}
\end{tcolorbox}
\caption{An example of hallucination by ChatGPT-4o in response to a hardware security query.
While \textit{SILVER} \cite{knichel2020silver} exists in the literature, its described methodology and goals here are entirely incorrect.
\textit{RTL-PSC} \cite{rtlpsc} is a real framework but is partially misrepresented, and \textit{SCARF} is entirely fictitious. The red-highlighted lines indicate the incorrect information.}
\label{fig1}
\vspace{-4mm}
\end{figure}

A natural candidate to provide such guidance is a general-purpose large language model (LLM) such as \textit{ChatGPT} or \textit{Gemini}. However, applying these models directly to hardware security poses serious risks. Despite their impressive language capabilities, these models have two fundamental limitations that make them unsuitable for high-assurance domains. The first is \textbf{(1)} \textit{hallucination}, the tendency to produce factually incorrect or fabricated content. The second is their \textbf{(2)} \textit{static knowledge base}: \new{although modern LLMs increasingly support web-based retrieval to access recent information, these capabilities remain shallow and unstructured for specialized technical domains.} General-purpose LLMs still cannot reliably incorporate newly discovered attacks, emerging threats, or evolving verification practices \new{with the depth and precision required in hardware security}. Relying on a tool that may confidently invent ``facts'' and lacks awareness of current security practices is unacceptable in a domain where correctness is the primary objective, as illustrated in Figure \ref{fig1}. These limitations raise our second research question:

\textbf{\emph{\underline{RQ2}}: How can a conversational assistant for hardware security verification be designed to deliver trustworthy, up-to-date, and verifiable guidance while avoiding the risks of hallucination and knowledge staleness?}

Despite the limitations of general-purpose LLMs, a \textit{conversational AI assistant} remains a promising direction for addressing RQ1, provided it is designed with safeguards to address RQ2. Unlike static documentation, a conversational system can dynamically adapt to the evolving context of the engineer's tasks, allowing them to ask specific questions, clarify uncertainties, and explore alternative strategies as their verification work progresses. By shifting the effort from manual information gathering to focused question--answer exchanges, such an assistant can reduce the cognitive burden and allow engineers to devote their expertise to high-level security reasoning rather than low-level information search. The effectiveness of this interaction paradigm has already been demonstrated in related hardware design and verification contexts: for example, \textit{ChatIoT} \cite{chatiot}, \textit{ORAssistant} \cite{orassistant}, \textit{EDA-Copilot} \cite{edacopilot}, and \textit{ChatEDA} \cite{chateda} have shown how conversational agents can enhance engineers' productivity, but none have been designed to address the unique challenges of hardware security verification. \new{Notably, \textit{ChatEDA} focuses on autonomous EDA script generation without domain-specific security knowledge, and \textit{ORAssistant} provides RAG-based assistance for the OpenROAD flow without verification tool execution. In contrast, \textit{VeriChat} uniquely combines security-specialized RAG with direct EDA tool integration, enabling both domain-grounded guidance and concrete verification actions within a single conversational workflow.} More recently, LLM- and agent-based approaches have been explored for a variety of hardware security tasks \cite{hasan2026ai, saha2025sv, saha2024llm, saha2024empowering}, including vulnerability analysis \cite{tarek2025socurellm,malls,11563377}, security asset identification \cite{lasset}, formal security verification \cite{tarek2026assertain, ankireddy2025lasso,10.1145/3764934}, and test plan generation \cite{saha2025threatlens}. However, these efforts are designed for specific verification tasks rather than to provide comprehensive conversational support throughout the hardware security verification lifecycle. 

To address these challenges, we propose \textbf{\textit{VeriChat}} \footnote{The benchmark, case studies, and evaluation results used in this work are released at \url{https://bit.ly/3QPBiWr}.}, a domain-specialized conversational assistant designed to support, rather than replace, existing hardware security verification practices. Through interactive dialogue, it provides security-oriented guidance across all stages of the verification flow, from clarifying security concepts and reasoning about threat relevance to exploring verification methodologies, troubleshooting issues, and devising countermeasures. Crucially, it addresses the shortcomings of general-purpose LLMs by grounding its responses in curated, up-to-date security knowledge and applying strict validation strategies to prevent hallucinations. The key contributions of \textit{VeriChat} are as follows:
\begin{enumerate}
    \item \textit{\textbf{Multi-Agent Conversational Framework:}} We present \textit{VeriChat}, the first multi-agent chatbot designed to provide security-focused guidance throughout the hardware security verification flow via interactive conversation.
    \item \textit{\textbf{Comprehensive Domain Knowledge Base:}} We construct a large, curated, topic-partitioned database of 28K+ hardware security verification research papers and resources, enabling precise and context-rich information retrieval.
    \item \new{\textit{\textbf{EDA Tool Integration:}} We integrate open-source EDA tools (Icarus Verilog, Yosys, SymbiYosys) into the conversational workflow, enabling \textit{VeriChat} to perform syntax checking, synthesis analysis, simulation, and formal verification on user-provided RTL designs.}
    \item  \textit{\textbf{Comprehensive Evaluation Framework:}} We design a multi-faceted evaluation framework that rigorously assesses \textit{VeriChat} at both the component and system levels to ensure its reliability and trustworthiness.
\end{enumerate}

\section{Proposed Methodology}
\begin{figure}[t]
\centering
\includegraphics[scale=.6]{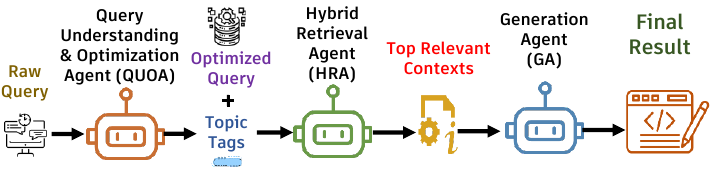}
\caption{Overview of \textit{VeriChat} Framework.}
\label{fig2}
\vspace{-4mm}
\end{figure}

\textit{VeriChat}, shown in Figure \ref{fig2}, uses a sequential multi-agent pipeline to ensure accurate answers. The Query Understanding and Optimization Agent (QUOA) first interprets and refines the user's query, clarifying its intent and tagging its topic. The Hybrid Retrieval Agent (HRA) then searches curated knowledge bases and the web, merging and reranking results into a unified evidence set. Finally, the Generation Agent (GA) produces a response strictly based on this verified context, reducing hallucinations and ensuring factual reliability. \new{When the user's query involves a design file and requires verification actions, \textit{VeriChat} routes the request through an integrated tool pipeline (Section~\ref{sec:eda}) that performs syntax checking, synthesis analysis, simulation, and formal verification on the provided RTL.}

\subsection{Query Understanding and Optimization Agent (QUOA)}
\begin{figure}[ht]
\centering
\includegraphics[scale=.065]{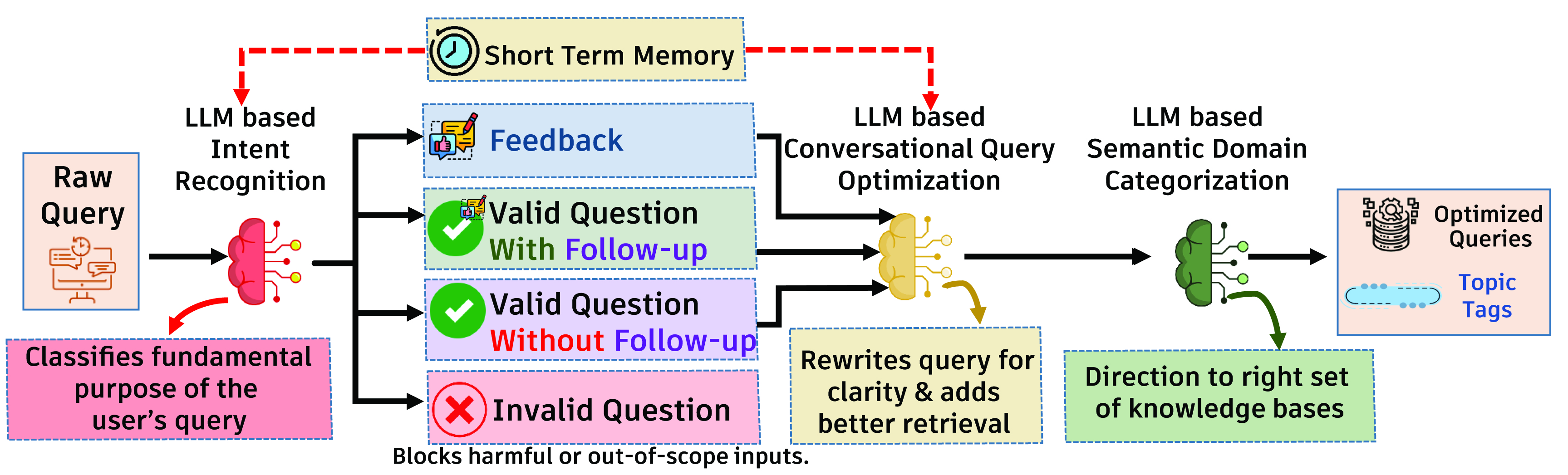}
\caption{Overview of QUOA.}

\label{fig3}
\end{figure}

The primary objective of QUOA is to act as a sophisticated gatekeeper and interpreter, transforming raw, often ambiguous user input into a validated, optimized, and actionable data packet. As shown in Figure \ref{fig3}, its operation is divided into three sequential core functions.

\subsubsection{Intent Recognition}
 QUOA first performs a triage step that classifies the user's input by analyzing both the raw query and the previous conversation turn. Using a few-shot prompted LLM as a multi-class classifier, it assigns the input to one of four classes: \textit{Invalid Question} (to block harmful or out-of-scope inputs), \textit{Feedback} (purely conversational remarks), \textit{Valid Question With Follow-Up} (new query referencing prior context), and \textit{Valid Question Without Follow-Up} (standalone information-seeking query). \new{A secondary intent classifier determines whether the query requires verification tool execution (e.g., ``Run a security analysis on this design''), routing such requests to the pipeline described in Section~\ref{sec:eda}.} This classification ensures system safety and tailors the subsequent workflow, allowing only non-\textit{Invalid Question} inputs to proceed to the core logic.

\subsubsection{Conversational Query Optimization}
In multi-turn dialogues, user queries often become progressively more ambiguous and context-dependent due to the use of conversational shortcuts such as coreference (referencing previously mentioned entities by pronouns) \cite{kostric2024surprisingly} and ellipsis (omitting information recoverable from context) \cite{su2019improving}. While these strategies are natural in conversation, they produce queries that are underspecified or incomprehensible when taken out of context, making them unsuitable for retrieval systems that require explicit, standalone formulations. Even single-turn queries can suffer from vagueness, using informal or overly broad phrasing that fails to capture the user's precise technical intent.

To address this, we use an LLM guided by a domain-aware prompt to produce two optimized outputs: one for dense retrieval and another for web search. The prompt handles both single- and multi-turn inputs and provides domain-specific rewriting rules and examples to ensure semantic clarity. For dense retrieval, the LLM generates a self-contained, terminology-aligned query suitable for semantic embedding (Figure \ref{fig:query_optimization}); for web search, it produces a keyword-focused version aligned with technical phrasing. If the query is classified as \textit{Valid Question With Follow-Up}, short-term memory is incorporated during the rewriting.

\begin{figure}[h]
\centering
\begin{tcolorbox}[
  width=\columnwidth,
  colback=yellow!5!white,
  colframe=gray!70!black,
  sharp corners,
  boxrule=1pt,
  left=1pt,right=1pt,top=1pt,bottom=1pt
]
\scriptsize
\textbf{\textcolor{red}{1st turn:}}
\textbf{``original query''}: ``I have a Register-Transfer Level (RTL) implementation of an AES-128 core written in Verilog. I need to perform a comprehensive pre-silicon security assessment to identify vulnerabilities to power side-channel attacks. Specifically, I am looking for a detailed overview of established frameworks that can be used for power side-channel leakage assessment.''
\textbf{\textcolor{teal}{``dense retriever optimized query''}}: ``What are the established frameworks for conducting a comprehensive pre-silicon security assessment of power side-channel leakage vulnerabilities in a Register-Transfer Level (RTL) implementation of an AES-128 core written in Verilog?''
\textbf{\textcolor{blue}{``google search optimized query''}}: ``Frameworks for power side-channel leakage assessment AES-128 RTL Verilog.''

\textbf{\textcolor{red}{2nd turn:}}
\textbf{``original query''}: ``Can you break down 2nd framework more?''
\textbf{\textcolor{teal}{``dense retriever optimized query''}}: ``Provide a detailed breakdown of the SCAR framework, which utilizes Graph Neural Networks for pre-silicon power side-channel analysis in RTL designs, focusing on its methodology and application in detecting vulnerabilities in cryptographic hardware.''
\textbf{\textcolor{blue}{``google search optimized query''}}: ``SCAR framework Graph Neural Networks RTL power side-channel analysis''
\end{tcolorbox}
\caption{Two-turn query optimization examples.}
\label{fig:query_optimization}
\vspace{-2mm}
\end{figure}

\subsubsection{Semantic Domain Categorization}
Naive RAG systems \cite{query} often rely on a single large, heterogeneous knowledge base. When queries target narrow subdomains (\textit{e.g.}, ``RTL-level side-channel leakage assessment for cryptographic cores''), searching such broad databases frequently retrieves semantically similar but topically irrelevant content (\textit{e.g.}, general cryptography or chip architecture). This distracts the LLM and degrades generation quality. To address this, \textit{VeriChat} uses a multi-retriever architecture with multiple specialized, topically coherent vector databases (\textit{e.g.}, \textit{pre-silicon side-channel leakage assessment}, \textit{RTL-level security}).

As the final QUOA step, the optimized query is classified by an LLM into one or more relevant topic labels from a predefined set aligned with these knowledge bases. This multi-label classification produces topic tags that are passed with the optimized queries to the next agent.

\begin{figure}[htbp]
\centering
\includegraphics[scale=.07]{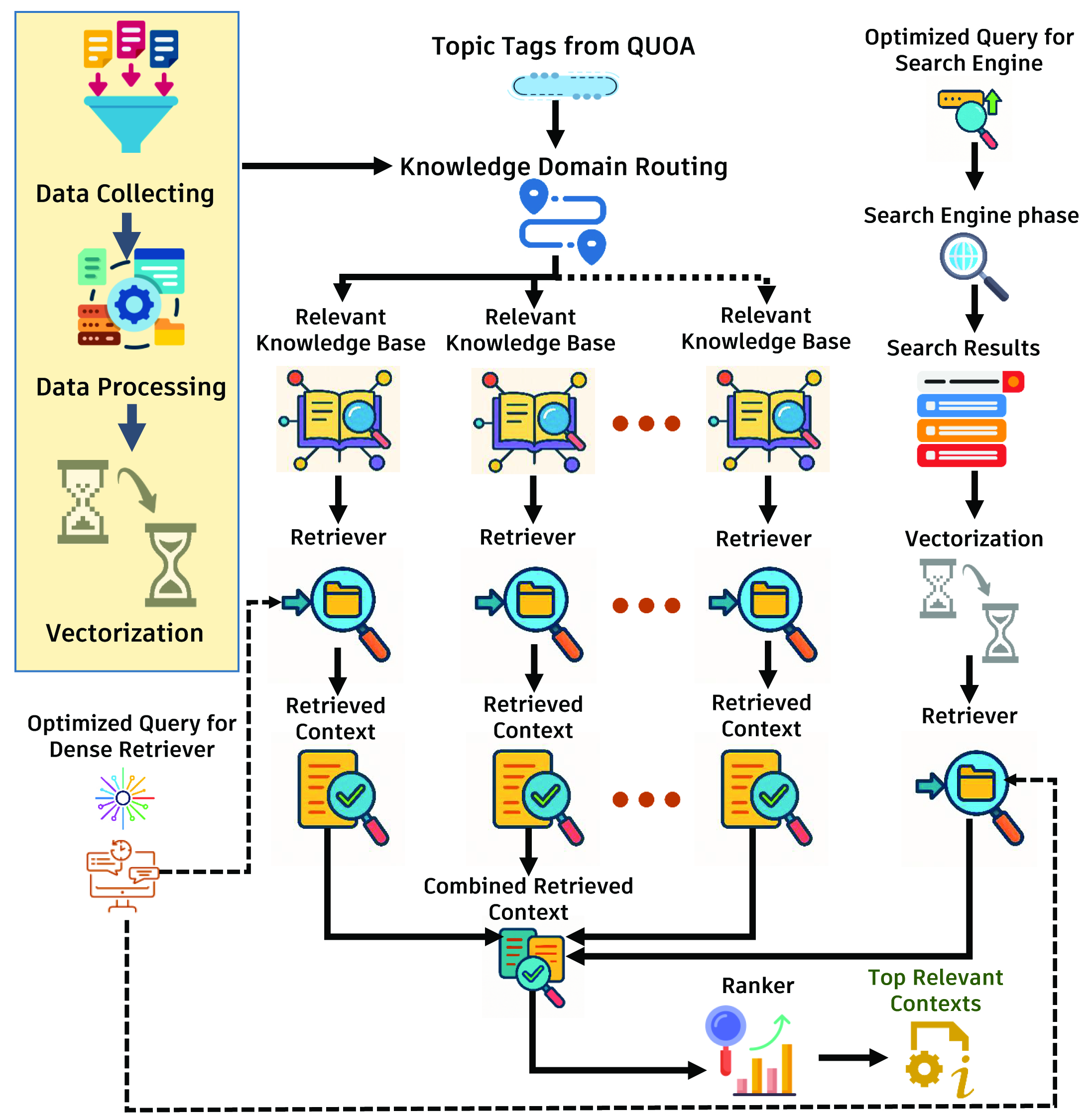}
\caption{Overview of Hybrid Retrieval Agent.}
\label{fig4}
\vspace{-4mm}
\end{figure}

\subsection{Hybrid Retrieval Agent (HRA)}
The HRA compiles a comprehensive and reliable evidence set for the QUOA's optimized queries using a hybrid strategy that combines high-precision retrieval from curated domain-specific knowledge bases with broad coverage from real-time web sources, ensuring the evidence is both accurate and up-to-date (Figure \ref{fig4}).

\subsubsection{Construction of Internal Knowledge Base}
The HRA's internal retrieval capability is built on a comprehensive, domain-specific knowledge base. We first defined a hierarchical taxonomy of 225 topics covering the hardware security verification domain. These topics serve a dual role: they structure the knowledge base and form the same set of classes used by the QUOA for Semantic Domain Categorization, ensuring a direct mapping between query classification and targeted retrieval. Using these topics as search queries, a custom Python-based web scraper collected high-quality, open-source academic papers and articles. To maintain relevance, each document was manually validated against its assigned topic before inclusion. The raw corpus then underwent a rigorous curation pipeline involving de-duplication, cleaning, and preprocessing, yielding 28,221 curated articles totaling about 61 GB. These documents were grouped by topic, segmented into overlapping text chunks, and transformed into high-dimensional vector embeddings using state-of-the-art models. The resulting embeddings were indexed into separate topic-specific vector databases, enabling the HRA's efficient multi-retriever search strategy.

\subsubsection{Parallel Retrieval Mechanisms}
Upon receiving the optimized query and its associated topic metadata from the QUOA, the HRA initiates multiple retrievers that operate in parallel:
\begin{description}[leftmargin=*, labelindent=0.5mm, labelsep=1.5mm]
\item[$\bullet$] \textit{Targeted Internal Retrievers:} For each topic tag identified by the QUOA, a dedicated retriever is activated. This retriever targets only the specific vector database corresponding to its assigned topic (e.g., \textit{RISC-V}, \textit{Spectre\_Vulnerability}). Within this partitioned database, it employs vector search to identify the top-k document chunks whose vector embeddings have the highest cosine similarity to the embedding of the optimized query. This multi-retriever approach on partitioned data is the primary reason for the system's speed and its ability to retrieve highly precise, domain-specific information.
\item[$\bullet$] \textit{Real-time External Retriever:} To account for information not present in the static knowledge bases, a separate retriever queries the open web. The optimized query is submitted to a Google Search API, \new{which returns results ranked by Google's proprietary relevance algorithm incorporating query-document similarity, page authority, and freshness.} \new{A} Web Scraper processes the content of the top-ranked search results. This transiently collected web content is then chunked and converted on-the-fly into a temporary vector database. A vector search is then performed on this ``in-memory'' database to retrieve the most relevant chunks, ensuring that the system's knowledge is not stale.
\end{description}
In both cases, a chunk size of 1,000 characters with an overlap of 200 characters is used to preserve context across segments. The embedding model used during vectorization is \textit{text-embedding-ada-002} from OpenAI, and Facebook AI Similarity Search (FAISS) is employed \new{with cosine similarity as the distance metric to rank all chunks in the target vector store and select the top \mbox{$k{=}20$} document chunks per retriever, where selection is based strictly on descending similarity score}.

\subsubsection{Fusion and Re-ranking}
Since relevance scores from different retrievers are not directly comparable, the HRA fuses them using a modified reciprocal rank fusion (RRF) \cite{reciprocal} with source-specific weighting. The weighted RRF score for each document chunk $c$ is computed as:
\begin{equation}
\text{WeightedRRF}_{\text{score}}(c) = \sum_{i=1}^{N} w_i \cdot \frac{1}{k + \text{rank}_i(c)}
\end{equation}
where $w_i$ is the weight assigned to the $i$-th source retriever, $\text{rank}_i(c)$ is the chunk's position in the $i$-th ranked list, and $k$ is a damping constant (set to 60). For instance, the internal retriever corresponding to the primary topic tag and the external Google search retriever are assigned equal priority ($w_i=0.4$), while retrievers for additional internal topics receive proportionally smaller weights.

The resulting comprehensive and re-ranked list of all retrieved chunks is stored in a session-specific memory. From this list, the top $M$ chunks (in our case $M$ = 100 at most) are selected and passed to the Generation Agent as the final context for generating the answer.
\vspace{-2mm}
\subsection{Generation Agent (GA)}
\label{sec:eda}
The GA is the final stage of the framework, responsible for both verifying the retrieved evidence and composing the user-facing response. Its objective is to first act as a quality assurance gate and then to synthesize the verified information into a fluent, well-articulated, and faithfully grounded answer.

\paragraph{Sufficiency Check}
Before attempting to generate a response, the GA must first ensure that the provided context is sufficient. The agent first analyzes the top-k chunks of context from the HRA to determine if they are sufficient to answer the query. If the initial context is deemed insufficient, the agent accesses the session-specific short-term memory, which contains the complete, unfiltered list of all documents retrieved by the HRA, and attempts to find more relevant information from this larger pool. If the context remains insufficient, the agent is designed to decline elaboration, thereby avoiding a speculative answer gracefully.

\paragraph{Prompt-Guided Faithful Generation}
Once the contexts seem sufficient, the GA proceeds to synthesize the information by creating an internal ``answer plan'' that identifies key themes and establishes a logical flow. Using this plan, the agent generates the final response. A detailed prompt meticulously controls this generation process, enforcing exclusive grounding (using only verified context), dynamic formatting (choosing the best format for clarity), rigorous attribution (including complete, IEEE-style citations), and a final self-scrutiny step to ensure the output is fully aligned with all constraints. The GA acts as a self-verifying synthesizer, transforming a raw knowledge corpus into a trustworthy and effective final response.

\paragraph{\new{EDA Tool Integration}}
\new{Prior conversational assistants for hardware design cannot interact with actual verification toolchains \cite{chateda, orassistant}. To address this, \textit{VeriChat} integrates open-source EDA tools into its workflow. When the QUOA classifies a query as an} \textit{eda\_tool\_execution} \new{intent, the system triggers a four-stage pipeline: (1)~\textit{Syntax Check} using Icarus Verilog; (2)~\textit{Synthesis Analysis} using Yosys to extract structural metrics such as flip-flop counts that may reveal suspicious overhead; (3)~\textit{Simulation} using Icarus Verilog and VVP with an LLM-generated targeted testbench; and (4)~\textit{Formal Verification} using SymbiYosys with bounded model checking via the Z3 SMT solver. At each stage, the LLM interprets raw tool output in the context of RAG-retrieved security knowledge, mapping findings to CWE categories and known attack patterns. If any stage fails, the system automatically refines its generated artifacts and retries up to three times using the failure logs to guide correction.}

\begin{table*}[t]
\centering
\scriptsize
\setlength{\tabcolsep}{4pt}
\renewcommand{\arraystretch}{0.92}
\caption{Qualitative comparison between \textit{VeriChat} and \textit{ChatGPT-5} responses on representative security verification tasks (full transcripts in the results repository under \texttt{CaseStudies/}).}
\label{tab:verichat_comparison}
\begin{tabularx}{\textwidth}{p{0.35\textwidth} p{0.11\textwidth} p{0.49\textwidth}}
\toprule
\textbf{Query} & \textbf{Related Task} & \textbf{Reasoning behind \textit{VeriChat's} superiority} \\
\midrule
\$\textit{Security Requirement Specification}\$
\textit{I prepared this Security Requirement Document for the I2C protocol. Now I want to know if there is any room for improvement to include more comprehensive security requirements. Also, what are the current practices that I missed in my requirement document?}
& Security Requirements Specification & VeriChat delivered a superior analysis by structuring its feedback around your SRD's specific design components, making it an immediately actionable engineering review. It critically fortified its recommendations with verifiable sources like the Accellera IPSA standard and academic papers, providing an expert authority that ChatGPT-5's generic, unsourced checklist lacked. This blend of tailored feedback and sourced credibility makes its guidance more professional and trustworthy.\\
\midrule
\textit{I want to identify security assets from the RTL designs. I am trying to develop a flow for this. How to do this?}
& Security Asset Identification & VeriChat provides a detailed, expert-level methodology that is specific and actionable. It incorporates advanced concepts like threat modeling, concolic testing, and information flow tracking, and cites relevant academic references, lending its answer authority and verifiability that GPT-5 lacks. \\
\midrule
\textit{I have developed an asset list for an RTL design. Now, I want to connect these assets with relevant threat models. Unfortunately, the threat models are scattered around the literature, and they are slightly different in description from one another from one literature to another. Hence, I am finding it difficult to unify this with my asset list. Can you show me a structured way to connect assets with relevant threat models?}
& Threat Modeling & VeriChat's response is superior as it details an actionable security verification methodology aligned with engineering workflows, while ChatGPT-5 offers a more conceptual framework. VeriChat's prescriptive process directly addresses the core requirement by normalizing scattered threat data into a scalable meta-model based on RTL trust boundaries and lifecycle phases. This defines an end-to-end workflow that culminates in concrete engineering deliverables, including a unified threat register and verifiable SVA property packs. Ultimately, this provides a clear, auditable path to security sign-off, a critical advantage over the more generalized guidance. \\
\midrule
\textit{Can you show me a suitable cost function required for gray box fuzzing on my RISC-V processor design so that it can identify all access control-related vulnerabilities?}
& Verification Planning & VeriChat offers precise, mathematically-defined cost functions for concrete RISC-V vulnerabilities (like PMP bypasses and illegal CSR access), based on a verifiable academic paper (SoCFuzzer). ChatGPT-5 gives a generic template using abstract heuristics, which is less reliable for real-world fuzzing. \\
\midrule
\textit{My formal tool is not converging on a security assertion that I have written. What should I do now?}
& Debugging & VeriChat provides specialized, expert-level guidance with correct academic references. It explains that many security properties are hyperproperties, needing specific techniques. ChatGPT-5 offers generic advice that fails to address the unique challenges of security assertions. \\
\midrule
\textit{I am performing a power side-channel leakage assessment for an AES-128 design without countermeasures. The gate count is around 20k. For collecting 1M power traces at the RTL level, the tool required 30 minutes at my end with 8 CPU cores. Is my runtime compatible with the current practices in RT-level power side-channel leakage assessment?}
& Result Validation & VeriChat is fact-based and strategic, citing real academic sources. The ChatGPT-5 response, on the other hand, is fundamentally unreliable as it fabricates its citations to create a false veneer of authority. Beyond just reliability, VeriChat provides a more expert consultation by questioning the user's underlying methodology--correctly identifying that one million traces is excessive for noiseless RTL analysis. This elevates its guidance from a simple runtime validation to a strategic recommendation that aligns the user's work with established, efficient practices. \\
\bottomrule
\end{tabularx}
\end{table*}

\section{Experiments and Results}
To validate \textit{VeriChat}, we conducted a two-stage evaluation: a component-level analysis of each agent and an end-to-end assessment against baselines. All datasets, results, and scripts are available at \url{https://bit.ly/3QPBiWr}.

\begin{figure}[t]
\centering
\includegraphics[width=\columnwidth]{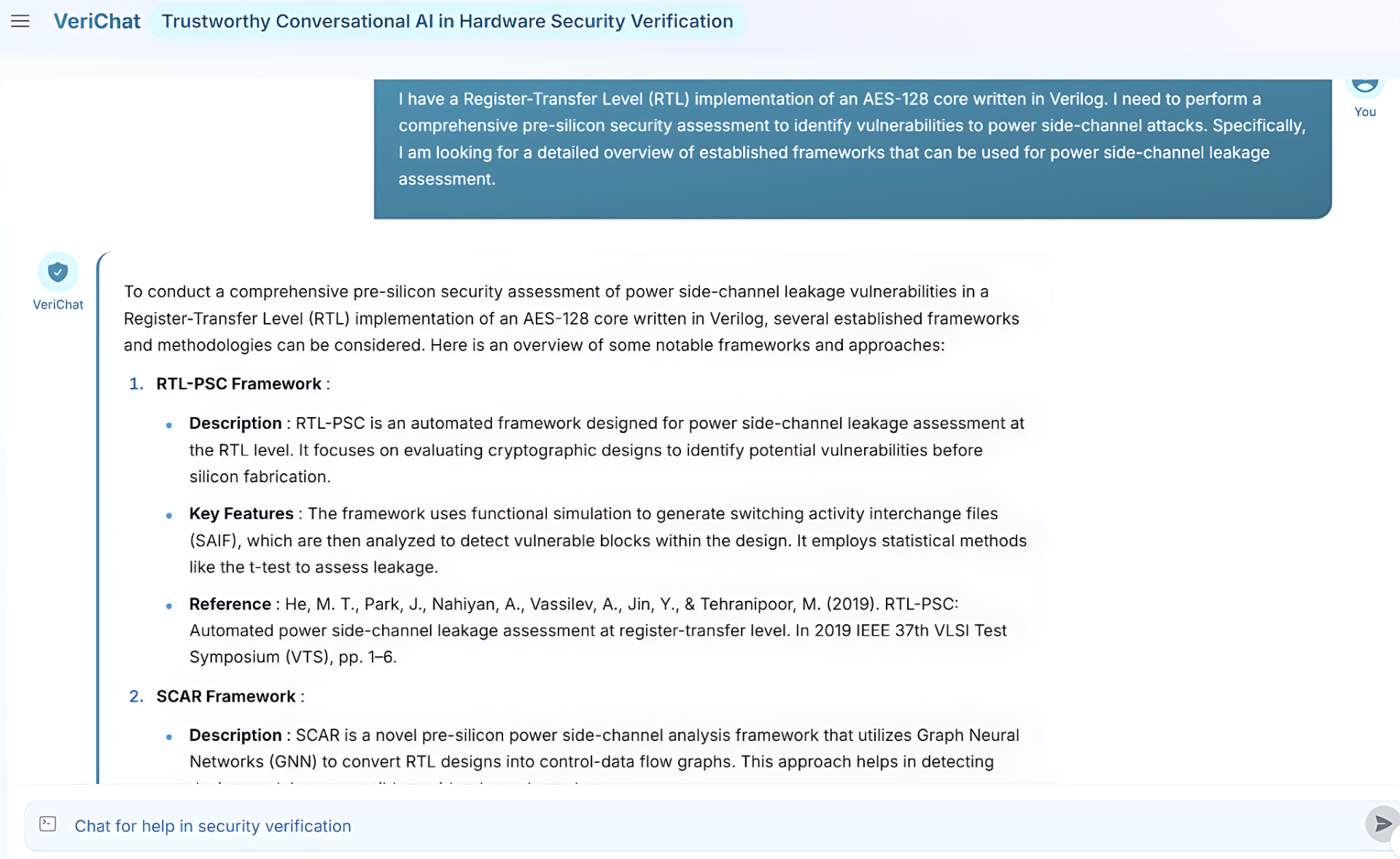}
\caption{Frontend interface of \textit{VeriChat} displaying a response to the same query shown in Figure \ref{fig1}. \textit{VeriChat} suggested RTL-PSC \cite{rtlpsc}, SCAR \cite{kanad_scar}, and other relevant frameworks with correct description.}
\label{fig:chat}
\vspace{-0.4em}
\end{figure}

\subsection{System Implementation}
\textit{VeriChat} (Figure \ref{fig:chat}) is implemented as a distributed system with a React.js frontend and a Flask backend connected via a RESTful API. The frontend offers an intuitive chat interface with components tailored for hardware security workflows, \new{including file upload for RTL designs and structured display of verification results}. The backend orchestrates the multi-agent framework and supports integration of any API-accessible model. In this work, \textit{GPT-4o} and \textit{GPT-5} serve as the primary generators. \new{For EDA tool execution, the backend interfaces with Icarus Verilog (native) and Yosys/SymbiYosys (via WSL) through secure subprocess calls with timeout enforcement and output truncation.} \new{Average end-to-end latency for Q\&A queries is roughly 15--25 seconds (dominated by LLM inference and multi-retriever search); tool execution adds 30--90 seconds depending on design complexity. Although the current implementation uses proprietary API\nobreakdash-accessible models, the modular architecture lets any API\nobreakdash-compatible LLM serve as a drop-in replacement, enabling future use of open-source alternatives.}

\subsection{Case Studies}
To demonstrate \textit{VeriChat}'s effectiveness across stages of hardware security verification, we conducted representative case studies comparing its responses with those of \textit{ChatGPT-5} (Table~\ref{tab:verichat_comparison}). The queries span key activities, including security requirement specification, asset identification, threat modeling, verification planning, debugging, and result validation. In each case, \textit{VeriChat} provided more structured, domain-specific, and evidence-backed guidance, while \textit{ChatGPT-5} produced generic or unsourced responses.

\subsubsection{EDA-Integrated Case Study: Hardware Trojan Detection in a Third-Party AES S-Box IP}
We demonstrate \textit{VeriChat}'s end-to-end workflow through a realistic scenario in which a user with no prior knowledge of a third-party AES S-Box IP progressively uncovers a hidden hardware Trojan through natural conversation. The design under test contains a rare-trigger sequential Trojan (Listing~\ref{lst:trigger}): a three-state FSM activated by the byte sequence \texttt{0xDE}, \texttt{0xAD}, \texttt{0xBE}, which arms a payload that leaks the secret key through \texttt{status\_led} over eight clock cycles. Per the Trust-HUB taxonomy, this is a Type-I sequential rare-event trigger with a covert side-channel payload, mapping to CWE-1245, with an activation probability of $\frac{1}{2^{24}} \approx 6 \times 10^{-8}$ per valid cycle.

\begin{lstlisting}[style=prettyverilog, caption={Trojan trigger: rare 3-byte sequence detector FSM.}, label={lst:trigger}]
reg [1:0] trig_cnt;
always @(posedge clk or negedge rst_n) begin
  if (!rst_n)
    trig_cnt <= 2'd0;
  else if (trig_cnt == 2'd3)
    trig_cnt <= 2'd0; // single-cycle pulse
  else if (valid_in)
    case (trig_cnt)
      2'd0: trig_cnt <=
              (data_in==8'hDE) ? 2'd1 : 2'd0;
      2'd1: trig_cnt <=
              (data_in==8'hAD) ? 2'd2 : 2'd0;
      2'd2: trig_cnt <=
              (data_in==8'hBE) ? 2'd3 : 2'd0;
      default: trig_cnt <= 2'd0;
    endcase
end
\end{lstlisting}

The user begins by asking: \textit{``I received an AES S-Box module from a third-party IP vendor. What kind of security threats should I be worried about?''} The QUOA classifies this as a \textit{security\_q\_a} intent, optimizes the query, and routes it to three topic-specific vector stores (\textit{hardware\_trojan}, \textit{supply\_chain\_attacks}, \textit{side\_channel\_attacks}). The HRA fuses retrieved chunks via weighted RRF alongside real-time web results, and the GA synthesizes a response covering seven threat categories with actionable mitigations grounded in retrieved academic references:

\begin{tcolorbox}[colback=green!3!white, colframe=green!50!black, boxrule=0.6pt, left=2pt, right=2pt, top=2pt, bottom=2pt, title={\scriptsize \textbf{VeriChat Response (excerpt)}}]
\scriptsize
\textbf{Key security threats for a third-party AES S-Box IP:}\\[1pt]
\textbf{1. Malicious modifications} (hardware Trojans, backdoors): Hidden triggers in FSMs that enable bypass or degrade security; can evade functional verification. \textit{Mitigation:} fuzzing/pen-testing of RTL, evolutionary search for rare trigger sequences.\\[1pt]
\textbf{2. Side-channel leakage} (power, EM, timing): Data-dependent switching may leak key information. \textit{Mitigation:} TVLA-style fixed-vs-random tests, constant-time enforcement.\\[1pt]
\textbf{3. Fault-injection susceptibility:} Glitches enable differential fault analysis for key recovery.
\textbf{4. DFT/scan-chain exposure.} \textbf{5. Undocumented debug interfaces.} \textbf{6. Non-compliance with FIPS-197.} \textbf{7. Logic locking/obfuscation} affecting availability.\\[1pt]
\textit{[13 academic references cited]}
\end{tcolorbox}

The user then follows up: \textit{``How would I detect if there is a Trojan in it? Can you actually check my design?''} The QUOA detects this as a \textit{Valid Question With Follow-Up} and incorporates short-term memory from the previous turn into the query optimization, enriching the rewritten query with the AES S-Box and Trojan context already established. The HRA retrieves additional content on detection methods: formal verification, coverage-guided fuzzing, structural analysis, side-channel monitoring, and golden model comparison. The GA synthesizes these strategies and, recognizing the user's request for active analysis, informs the user that \textit{VeriChat} can execute its integrated EDA tool pipeline on an uploaded design file.

The user uploads the AES S-Box Verilog file and requests: \textit{``Run a full security analysis on this design.''} This time, the QUOA classifies the query as an \textit{eda\_tool\_execution} intent, triggering the four-stage pipeline described in Section~\ref{sec:eda}. In Stage~1, the syntax check reports zero errors and zero warnings; the interpretation notes that syntactic correctness alone does not preclude semantic vulnerabilities. In Stage~2, synthesis creates registers for \texttt{data\_out}, \texttt{valid\_out}, and internal pipeline signals, but also reveals 7 surplus flip-flops absent from a pure S-Box: \texttt{trig\_cnt[1:0]} (2), \texttt{trojan\_armed} (1), \texttt{leak\_bit\_cnt[2:0]} (3), and \texttt{status\_led} (1), consistent with Trust-HUB Trojan benchmarks. Augmented by retrieved CWE knowledge, \textit{VeriChat} flags this as consistent with CWE-1245 and identifies \texttt{status\_led} as structurally decoupled from the core S-Box datapath. In Stage~3, the system generates a targeted testbench (Listing~\ref{lst:tb}) that applies random functional vectors, a near-miss sequence, and then the exact trigger sequence with a known key.

\begin{lstlisting}[style=prettyverilog, float=tbp, floatplacement=tbp, caption={LLM-generated targeted testbench (excerpt).}, label={lst:tb}]
// Normal operation: random vectors avoiding
// accidental trigger bytes
for (i = 0; i < 12; i = i + 1) begin
  r = $random;
  if (r==8'hDE||r==8'hAD||r==8'hBE)
    r = r ^ 8'h5A;
  send_valid_byte(r);
end
// Near-miss: should NOT trigger Trojan
send_valid_byte(8'hDE);
send_valid_byte(8'hAD);
send_valid_byte(8'hBF); // wrong third byte
// Targeted Trojan trigger
key_in = 8'hA6; // 0b1010_0110
send_valid_byte(8'hDE);
send_valid_byte(8'hAD);
send_valid_byte(8'hBE);
// Monitor key leakage on status_led
leaked = 8'h00;
for (k = 0; k < 8; k = k + 1) begin
  @(posedge clk);
  leaked[k] = status_led;
  $display("[%0t] LEAK: bit%0d=%0b",
           $time, k, status_led);
end
\end{lstlisting}

Simulation confirms Trojan activation at 535\,ns with key-dependent information leakage through \texttt{status\_led} over 8 cycles, while normal S-Box functionality remains correct throughout Phases 1--2. In Stage~4, \textit{VeriChat} generates SVA properties encoding security invariants and invokes bounded model checking with the Z3 SMT solver \cite{z3}. The confidentiality property (Listing~\ref{lst:sva}) asserts that \texttt{status\_led} must remain inactive at all times in a purely functional S-Box.

The formal engine reports \texttt{BMC failed!} at step~1, producing a counterexample trace that mathematically proves the confidentiality violation. The auto-retry mechanism was exercised during this stage: the first attempt included overly broad assertions on internal Trojan signals (e.g., \texttt{trojan\_armed}, \texttt{trig\_cnt}) that trivially failed, and the second attempt retained an integrity check on \texttt{data\_out} that also violated. The system automatically fed each failure log back to the LLM, which progressively refined the property set until the third attempt isolated the targeted confidentiality assertion. This progressive revelation across the conversation (clean syntax, suspicious synthesis overhead, confirmed simulation, and formally proven violation) mirrors established multi-layered hardware security assessment methodologies.
\begin{lstlisting}[style=prettyverilog, caption={LLM-generated SVA confidentiality property.}, label={lst:sva}]
// Confidentiality: after the DE->AD->BE
// sequence, status_led must stay inactive
// for 8 cycles (no key leakage permitted)
reg [3:0] fv_leak_window;
always @(posedge clk) begin
  if (!rst_n)
    fv_leak_window <= 4'd0;
  else if (fv_mon_pulse)
    fv_leak_window <= 4'd8;
  else if (fv_leak_window != 4'd0)
    fv_leak_window <= fv_leak_window - 4'd1;
end
always @(posedge clk) begin
  if (rst_n && fv_past_valid)
    if (fv_leak_window != 4'd0)
      assert(status_led == 1'b0);
end
\end{lstlisting}

\subsection{Evaluations}
\subsubsection{Benchmark}
To facilitate a comprehensive evaluation, we curated a new benchmark by engaging 25 experienced researchers in the domain of hardware security verification. Each expert contributed a set of six distinct queries on a topic of their choice, covering fact-checking, reasoning, scenario-based analysis, real-time queries, tool usage queries, and multiple-choice questions. This process yielded a diverse and challenging benchmark comprising 150 queries, which we subsequently used to assess the performance of \textit{VeriChat} and other baseline models.
\begin{figure*}[t]
  \centering
    \begin{subcaptionbox}{Context Recall and Precision Test\label{fig8:sub4}}[0.160\textwidth]
    {\includegraphics[width=\linewidth]{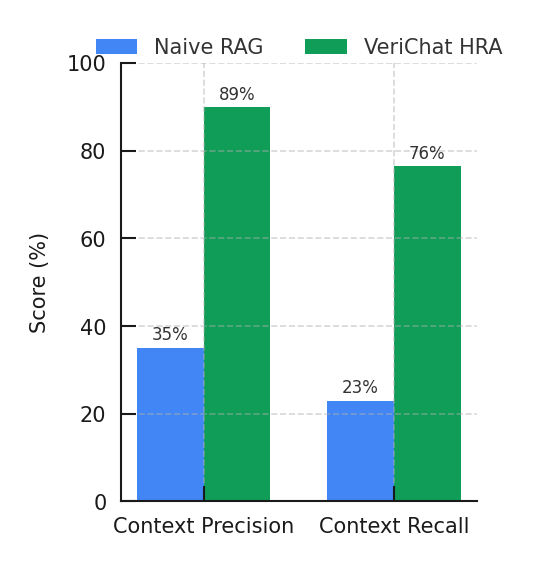}}
  \end{subcaptionbox}
  \hfill
  \begin{subcaptionbox}{Faithfulness Test\label{fig8:sub1}}[0.210\textwidth]
    {\includegraphics[width=\linewidth]{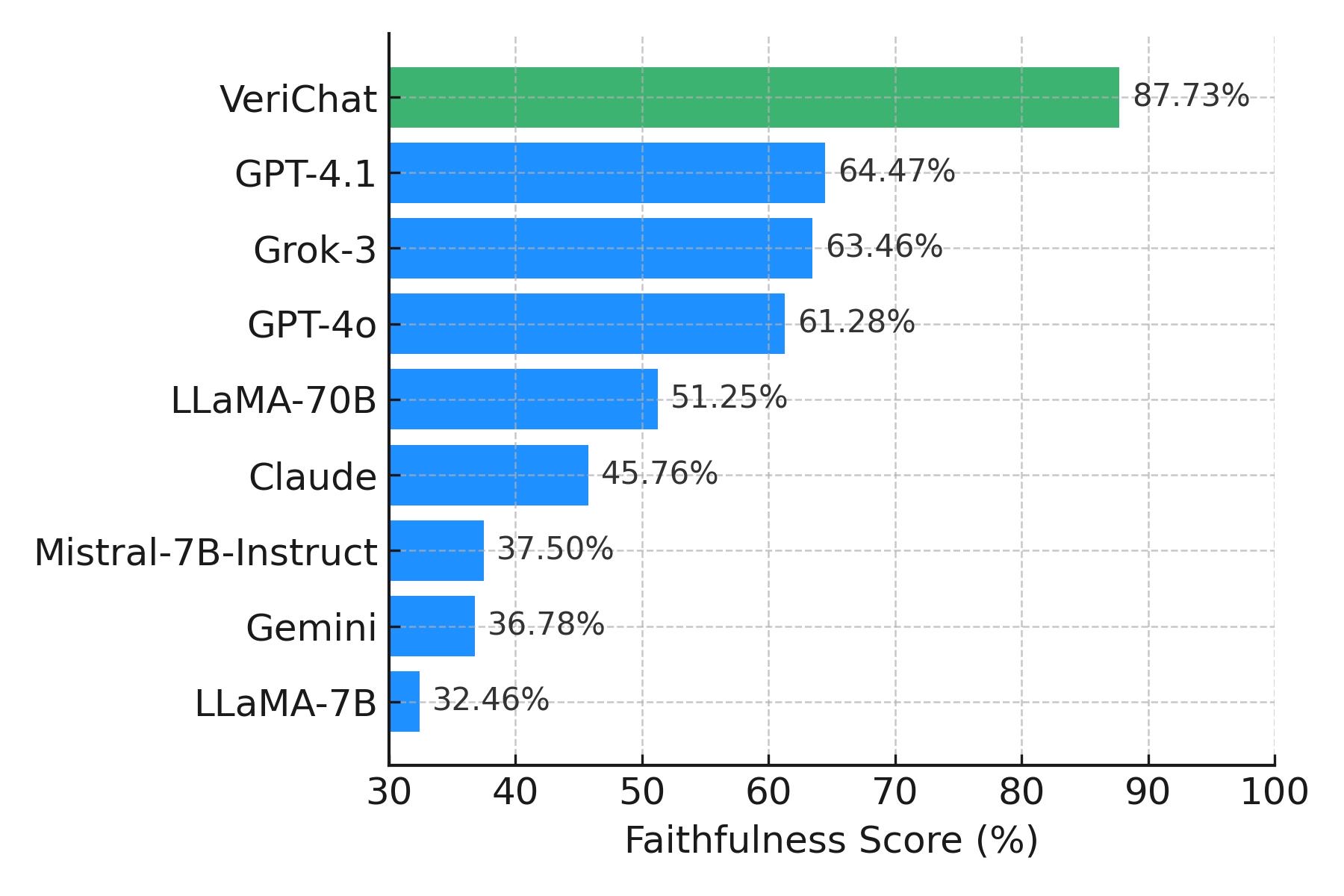}}
  \end{subcaptionbox}
  \hfill
  \begin{subcaptionbox}{Counterfactual Robustness Test\label{fig8:sub2}}[0.160\textwidth]
    {\includegraphics[width=\linewidth]{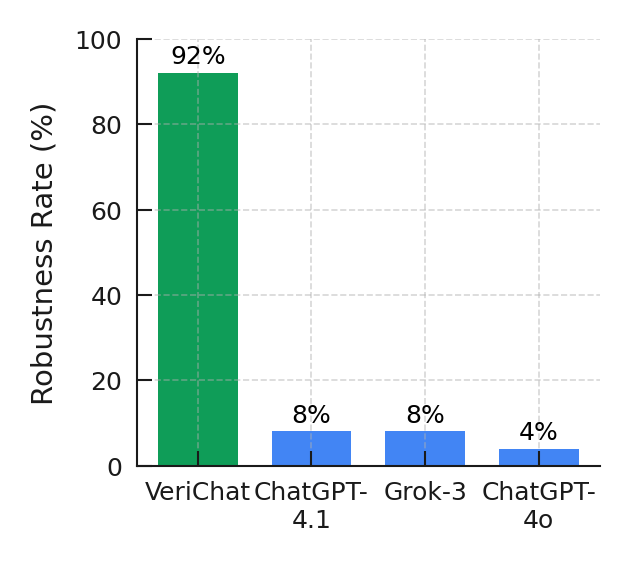}}
  \end{subcaptionbox}
  \hfill
  \begin{subcaptionbox}{User Preference Test\label{fig8:sub3}}[0.210\textwidth]
    {\includegraphics[width=\linewidth]{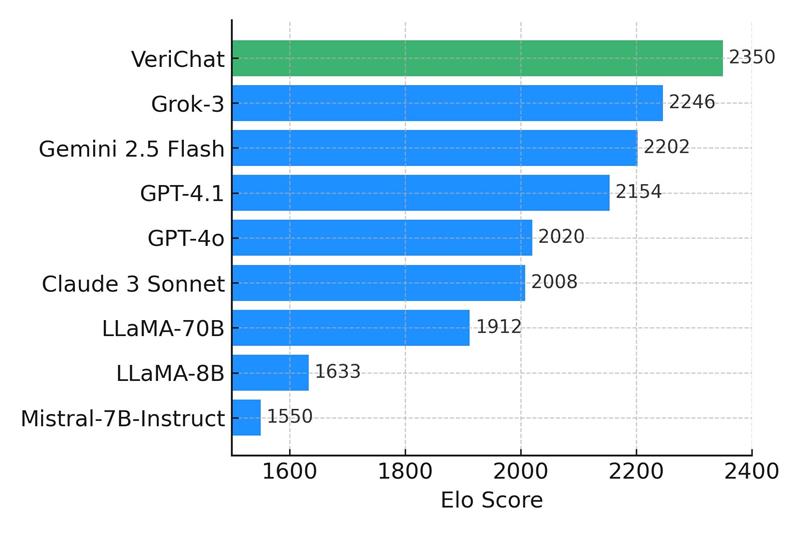}}
  \end{subcaptionbox}
  \hfill
  \caption{Performance of \textit{VeriChat} in different evaluations.}
  \label{fig:threeinrow}
  \vspace{-4mm}
\end{figure*}
\subsubsection{Component-Level Evaluation}

\paragraph{\underline{Test 1 \& 2}: Context Recall and Precision Evaluation}
We assessed \textit{VeriChat}'s Hybrid Retrieval Agent (HRA) using \textit{DeepEval} \cite{deepEval2024} on 150 question--answer samples, measuring both Context Recall (coverage of key facts) and Context Precision (ratio of relevant to retrieved content). An LLM-as-a-judge (\textit{GPT-4.1}) checked whether each gold-standard statement was supported by the retrieved context (recall) and whether each retrieved node was relevant to the query (precision). The metrics are defined as:
\begin{equation}
\text{Context Recall} = \left(\frac{\text{\# Attributable Statements}}{\text{\# Statements in Expected Output}}\right)
\end{equation}
\vspace{-3mm}
\begin{equation}
\text{Context Precision} = \frac{1}{R} \sum_{k=1}^{n} \left( \frac{R_k}{k} \cdot r_k \right)
\end{equation}
\noindent where $R$ is the total number of relevant nodes, $R_k$ is the number of relevant nodes up to position $k$, and $r_k$ is a binary indicator of relevance at position $k$.

\textit{VeriChat} achieved a Context Recall of 76.40\% and a Context Precision of 89.85\%, far surpassing the Naive RAG baseline (23.00\% and 20.31\% respectively) As shown in Figure~\ref{fig8:sub4}. This demonstrates that the QUOA-guided hybrid retrieval effectively gathers all essential information while minimizing irrelevant noise, unlike the baseline, which often misses key facts and includes off-topic content.

\paragraph{\underline{Test 3 \& 4}: Answer Relevancy and Prompt Alignment Evaluation}
We further evaluated \textit{VeriChat}'s Generation Agent on 150 prompts using \textit{DeepEval} \cite{deepEval2024}, measuring Answer Relevancy (topicality of the response) and Prompt Alignment (adherence to specified constraints). An LLM-as-a-judge (\textit{GPT-4.1}) scored each output for how well it addressed the user's query and followed the given instructions, including tone, persona, safety protocols, coherence, and formatting. The metrics are defined as:
\begin{equation}
\text{Answer Relevancy} = \frac{\text{\# of Relevant Statements}}{\text{Total \# of Statements}} \times 100\%
\end{equation}
\begin{equation}
\text{Prompt Alignment} = \frac{\text{\# of Instructions Followed}}{\text{Total \# of Instructions}} \times 100\%
\end{equation}
\textit{VeriChat} achieved 91.78\% in Answer Relevancy and 88.3\% in Prompt Alignment, indicating that its responses are not only factually grounded and on-topic but also reliably follow complex formatting, citation, and behavioral constraints.

\subsubsection{End-to-End System Evaluation}

\paragraph{\underline{Test 5}: Faithfulness Test}
This experiment asks: to what extent are a model's responses grounded in verifiable facts? We posed 40 hardware security questions to all nine models, anonymized the responses, and had a human domain expert conduct a blind review, decomposing each into discrete factual claims verified against ground-truth sources as ``correct'' or ``incorrect.'' The Faithfulness score is the ratio of correct to total claims:
\begin{equation}
    \text{Faithfulness Score} = \left( \frac{\text{\# correct claims}}{\text{\# total claims}} \right) \times 100\%
\end{equation}

The results shown in Figure \ref{fig8:sub1} reveal that \textit{VeriChat} is substantially more faithful to facts than its competitors, achieving a score of 87.73\%, which is over 23 percentage points higher than the next best model. This significant gap indicates the effectiveness of \textit{VeriChat}'s multi-agent, retrieval-first architecture in mitigating hallucination.

\paragraph{\underline{Test 6}: Counterfactual Robustness Test}
A trustworthy AI system must not only answer correctly but also reject user-introduced misinformation. To evaluate this, we designed an adversarial attack testing the Counterfactual Robustness of \textit{VeriChat} against other leading proprietary chatbots. This involved 25 carefully crafted ``hallucinated prompts'' that confidently embed fabricated concepts or factually incorrect premises and ask the model to elaborate. A sample prompt is shown below:
\begin{tcolorbox}[colback=yellow!5!white, colframe=yellow!50!black, boxrule=0.8pt, left=2pt, right=2pt, top=2pt, bottom=2pt]
\scriptsize
\textbf{Sample Hallucinated Prompt:}
A major innovation in anti-tamper technology is \textbf{Metamaterial Resonance Shielding (MRS)}.
This involves creating a smart mesh for ICs built from programmable metamaterials.
Any physical intrusion attempt (like micro-probing) changes the material's resonant frequency, triggering a response to zeroize critical data. Can you explain?
\end{tcolorbox}
The Counterfactual Robustness Rate can be expressed as:
\begin{equation}
    \text{Robustness Rate} = \left( \frac{\text{\# Successful Refusals}}{\text{\# Test Cases}} \right) \times 100\%
\end{equation}
\textit{VeriChat} achieved a Robustness Rate of 92\%, while monolithic proprietary models proved highly vulnerable (Figure \ref{fig8:sub2}). This robustness arises from \textit{VeriChat}'s strict reliance on retrieved evidence: when the HRA finds no support for a fabricated concept, the Generation Agent reports insufficient information instead of hallucinating.

\paragraph{\underline{Test 7}: User Preference Test}
We conducted a user preference evaluation using the Elo rating system, which is well-suited to evaluating chatbots through pairwise comparisons \cite{chatarenallm, chatarenahuman}. Following \cite{chatarenallm}, we designed a ``chatbot arena'' of 1,000 matches in which two randomly selected models from a pool of nine (\textit{VeriChat} and eight leading proprietary chatbots) responded to the same prompt. \textit{GPT-4o} served as an impartial judge, selecting the best of the two anonymous responses based on factual precision and supporting evidence.
Each match outcome updated the systems' Elo ratings using
\begin{equation}
    R'_A = R_A + K(S_A - E_A),
\end{equation}
where $S_A$ is the match score (1 win, 0.5 tie, 0 loss), $K$ a constant, and $E_A = (1 + 10^{(R_B - R_A)/400})^{-1}$ the expected score \cite{chatarenallm}. As shown in Figure~\ref{fig8:sub3}, \textit{VeriChat} achieved the highest Elo score (2350), over 100 points above its nearest competitor (\textit{Grok-3}) and nearly 200 above \textit{ChatGPT-4.1}, ranking as the most preferred chatbot for reliable, evidence-based responses.

\section{Conclusion}
\label{sec:conclusion}
We presented \textit{VeriChat}, a multi-agent conversational framework for hardware security verification that combines retrieval-augmented generation with integrated open-source EDA tools to deliver both domain-grounded guidance and concrete verification actions. In the hardware Trojan detection case study, it progresses from knowledge retrieval to tool-driven assessment (synthesis analysis, simulation, and formal verification) within a single multi-turn conversation. \textit{VeriChat} achieves 87.73\% faithfulness and 92\% counterfactual robustness, outperforming all eight baselines. Future work includes semi-automated knowledge base expansion, provenance tracking, and richer explainability such as citation-level attribution and confidence calibration.

\bibliographystyle{IEEEtran}
\bibliography{references}
\end{document}